\documentclass[reprint, groupedaddress, superscriptaddress, amsmath,amssymb, aps, pra]{revtex4-2}

\usepackage[colorlinks = true,]{hyperref}
\usepackage{float}
\usepackage{graphicx}
\usepackage{dcolumn}
\usepackage{bm}
\usepackage{xcolor}
\usepackage{braket}
\usepackage{soul}
\usepackage{comment}
\usepackage{tikz}
\usetikzlibrary{arrows.meta}

\begin{document}

\preprint{APS/123-QED}

\title{Engineering nonlinear activation functions for all-optical neural networks via quantum interference}

\author{Ruben Canora}
\altaffiliation{These authors contributed equally to this work}
\affiliation{Elmore Family School of Electrical and Computer Engineering, Purdue University, West Lafayette, Indiana 47907, USA}

\author{Xinzhe Xu}
\altaffiliation{These authors contributed equally to this work}
\affiliation{Elmore Family School of Electrical and Computer Engineering, Purdue University, West Lafayette, Indiana 47907, USA}

\author{Ziqi Niu}
\altaffiliation{These authors contributed equally to this work}
\affiliation{Elmore Family School of Electrical and Computer Engineering, Purdue University, West Lafayette, Indiana 47907, USA}

\author{Hadiseh Alaeian}
\email{halaeian@purdue.edu}
\affiliation{Elmore Family School of Electrical and Computer Engineering, Purdue University, West Lafayette, Indiana 47907, USA}
\affiliation{Department of Physics and Astronomy, Purdue University, West Lafayette, Indiana 47907, USA}
\affiliation{Purdue Quantum Science and Engineering Institute, Purdue University, West Lafayette, Indiana 47907, USA}

\author{Shengwang Du}
\email{dusw@purdue.edu}
\affiliation{Elmore Family School of Electrical and Computer Engineering, Purdue University, West Lafayette, Indiana 47907, USA}
\affiliation{Department of Physics and Astronomy, Purdue University, West Lafayette, Indiana 47907, USA}
\affiliation{Purdue Quantum Science and Engineering Institute, Purdue University, West Lafayette, Indiana 47907, USA}

\date{\today}

\begin{abstract}
\noindent All-optical neural networks (AONNs) promise transformative gains in speed and energy efficiency for artificial intelligence (AI) by leveraging the intrinsic parallelism and wave nature of light. However, their scalability has been fundamentally limited by the high power requirements of conventional nonlinear optical elements. Here, we present a low-power nonlinear activation scheme based on a three-level quantum system driven by dual laser fields. This platform introduces a two-channel nonlinear activation matrix with both self- and cross-nonlinearities, enabling true multi-input, multi-output optical processing. The system supports tunable activation behaviors, including sigmoid and ReLU functions, at ultralow power levels (17 $\mu$W per neuron). We validate our approach through theoretical modeling and experimental demonstration in rubidium vapor cells, showing the feasibility of scaling to deep AONNs with millions of neurons operating under 20 W of total optical power. Crucially, we also demonstrate the all-optical generation of gradient-like signals with backpropagation, paving the way for all optical training. These results mark a major advance toward scalable, high-speed, and energy-efficient optical AI hardware.
\end{abstract}

\maketitle

\section{introduction}\label{sec:introduction}
\begin{figure*}[t]
\centering
\includegraphics[width=0.95 \linewidth]{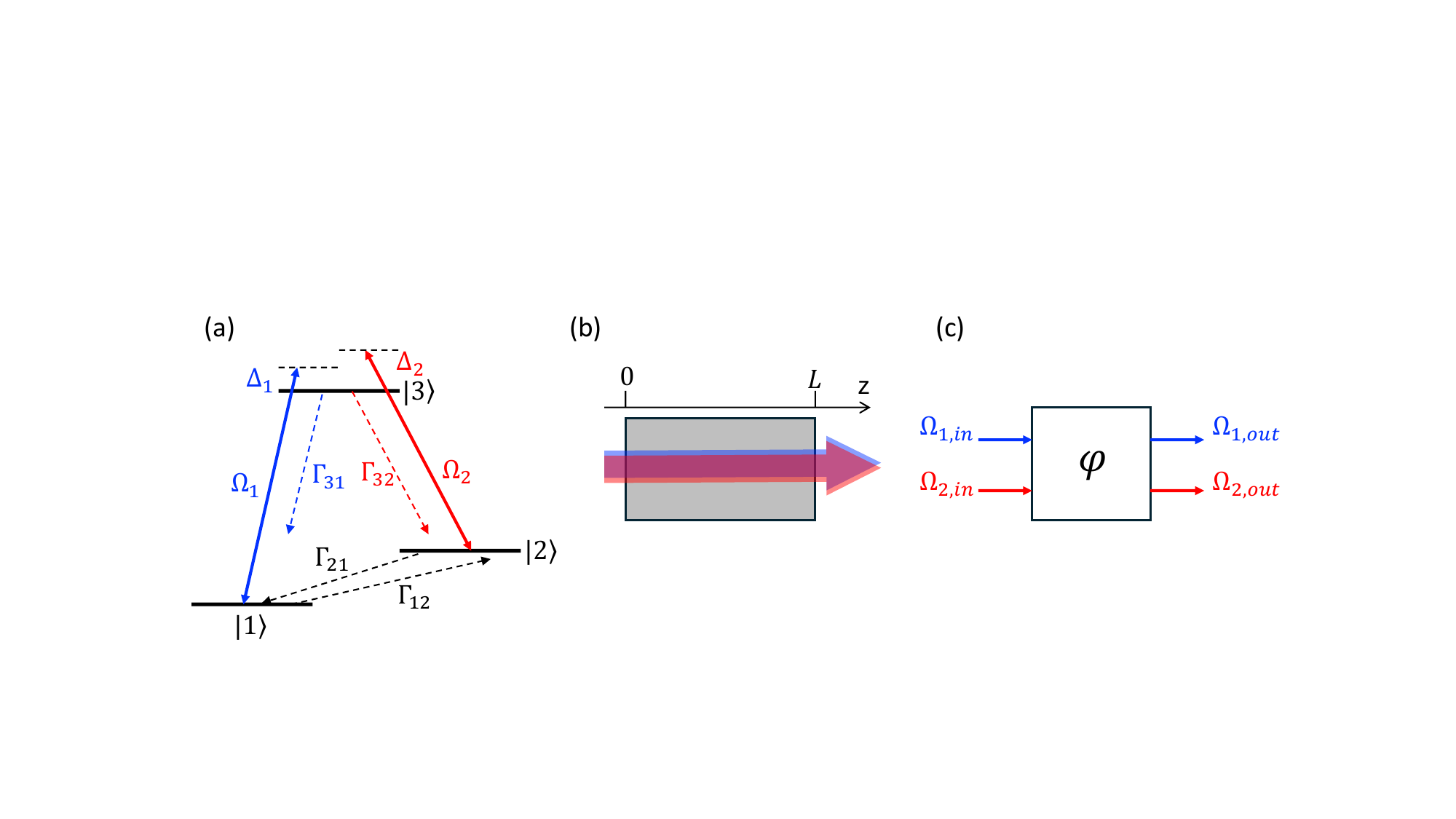}
\caption{\textbf{Schematic of the three-level nonlinear optical medium.} (a) The energy level diagram of a three-level quantum system with two driving lasers and decay mechanisms, which are considered in this work. (b) Optical setup showing the alignment of two laser beams and their propagation in the medium. (c) Simplified circuit diagram for the two-channel (2-input $\times$ 2-output) nonlinear activation function unit.}
\label{fig:fig01}
\end{figure*}

All-optical neural networks (AONNs) have emerged as a promising photonic computing paradigm for artificial intelligence (AI) applications, with potential in leveraging the wave nature of light to achieve intrinsic parallelism, ultrafast speed, and low energy consumption~\cite{NPhoton_Shastri2021, zuo2019all, PhysRevApplied.21.014028, 10.1063/5.0215752, light2024review, sui2020review, shariv1989all}. These attributes position AONNs as attractive alternatives to electronic platforms such as CPUs, GPUs~\cite{li2014large, she2019fast, keckler2011gpus}, TPUs, FPGAs~\cite{HAQ, zhang2016energy}, and ASICs~\cite{OLCEL, O3BNN}, particularly as the scaling limits of Moore's law become increasingly apparent~\cite{thompson2006moore, kish2002end}. Despite their advantages, the scalability of deep AONNs remains limited by the challenge of implementing nonlinear activation functions using purely optical mechanisms. These nonlinearities are essential for enabling deep learning but typically require high optical power, limiting the practicality of large-scale, energy-efficient AONN systems. While linear operations can be performed efficiently even at the few-photon level~\cite{NC13-2022}, the absence of robust low-power optical nonlinear elements restricts the depth and computational expressiveness of current architectures. As a result, most large-scale ONNs have focused primarily on linear transformations, often relying on hybrid optical-electronic designs to compensate for the missing nonlinearity~\cite{hamerly2019large, JGeorge2018, Spall:22}.

Several approaches have been explored to introduce optical nonlinearities into AONNs. For instance, phase-change materials have enabled chip-scale nonlinear functions, supporting demonstrations of only single-layer optical neurons~\cite{Nature_569_2019} -- The implementations remain limited in scalability. Another effort employed electromagnetically induced transparency (EIT) \cite{EIT-Harris, RevModPhys.77.633} in a laser-cooled atomic ensemble to realize nonlinear activation in a two-layer AONN~\cite{zuo2019all}. While conceptually significant, the large system footprint and operational complexity associated with laser cooling of atoms hindered its practical deployment. To date, experimental realizations of multilayer AONNs have typically been constrained to systems with only a few hundred hidden neurons~\cite{PRApplied-Du-2021}, underscoring the need for low-power, scalable solutions that can support deep architectures.

In this work, we present a quantum-optical strategy for implementing nonlinear activation functions using a three-level quantum medium driven by two laser fields. In contrast to conventional designs that implement single-channel scalar nonlinearities, our method enables a two-channel activation function that simultaneously exhibits self- and cross-nonlinear behavior. These nonlinearities arise from quantum interference between atomic transition pathways. Through a combination of theoretical modeling in both homogeneous and Doppler-broadened regimes, as well as experimental demonstration with a rubidium (Rb) vapor cell, we show that the system can emulate rectified linear unit (ReLU) and sigmoid activation functions at optical power levels as low as $17~\mu$W per neuron. This approach enables the design of deep AONNs containing millions of neurons operating with only less than 100 Watt-level laser sources. We further show that the scheme allows for backpropagation all-optical training. Our numerical simulations and experimental measurements confirm the scalability and energy efficiency of this architecture, offering a compelling pathway toward practical, high-speed, all-optical computing.

This article is organized as follows. Section~\ref{sec:lifetimebroadened} introduces the general formulation of a three-level quantum system, including the steady-state density matrix and light propagation equations, focusing on lifetime-broadened optical media. Section~\ref{sec:doppler-broadened} extends the analysis to Doppler-broadened media, confirmed by experimental measurements. Section~\ref{sec:conclusion} summarizes the work and suggests directions for future research.


\section{Lifetime-Broadened Medium}~\label{sec:lifetimebroadened}
%
Conventional non-linear optical processes typically operate at frequencies far detuned from resonances to avoid significant linear absorption ~\cite{boyd2003nonlinear, shen1984principles}. The $n^\textrm{th}$ order of nonlinear susceptibility can be expressed as $\chi^{(n)} \propto \frac{1}{\Delta_1+i\gamma_1} \frac{1}{\Delta_2+i\gamma_2} \cdots \frac{1}{\Delta_n+i\gamma_n}$, where $\Delta_m$ and $\gamma_m$ denote the frequency of detuning and dephasing of the $m$-photon process, respectively. To mitigate resonant absorption, $\Delta_m$ is generally required to be much larger than the absorption linewidth. Consequently, non-linear optics in solid-state materials require high light intensities due to their substantial absorption bandwidths, typically on the order of terahertz (THz)~\cite{boyd2003nonlinear}. 

On the other hand, both real and artificial atoms, such as quantum dots~\cite{michler2003single} and defects in solids~\cite{aharonovich2016solid}, with discrete energy levels, exhibit much narrower resonance linewidths, primarily determined by lifetime, making them promising candidates for achieving large nonlinear susceptibilities. Moreover, phenomena such as EIT~\cite{EIT-Harris, RevModPhys.77.633} and coherent population trapping~\cite{arimondo1996v} have demonstrated that resonant linear absorption can be suppressed or even eliminated through quantum interference in a multilevel system, thereby significantly improving nonlinear optical effects~\cite{schmidt1996giant, harris1998nonlinear}. While early demonstrations of AONNs using EIT have shown conceptual promise~\cite{zuo2019all}, their nonlinear activation functions have remained highly specialized, and a general framework for engineering diverse nonlinearities suited for machine learning remains lacking.

Here, we systematically investigate a three-level lifetime-broadened atomic medium and demonstrate the realization of both sigmoid and ReLU-like activation functions at ultralow optical power via quantum interference, surpassing the limitations of previous EIT work~\cite{zuo2019all}. Although our numerical analysis is based on $^{87}$Rb atoms, the conclusions are broadly applicable to other lifetime-broadened systems.

Figure~\ref{fig:fig01}(a) presents the energy level diagram of a three-level system, where $|1\rangle$ and $|2\rangle$ are two long-lived hyperfine ground states and $|3\rangle$ is an excited state. We define $\omega_{31} = (E_{3} - E_{1})/\hbar$ as the resonance frequency for the $\lvert1\rangle \rightarrow \lvert3\rangle$ transition and $\omega_{32} = (E_{3} - E_{2})/\hbar$ as the resonance frequency for the $\lvert2\rangle \rightarrow \lvert3\rangle$ transition. Optical field 1 (depicted in blue) couples the $|1\rangle \rightarrow |3\rangle$ transition with a Rabi frequency $\Omega_1$ and one-photon detuning $\Delta_1=
\omega_{1}-\omega_{31}$, where $\omega_{1}$ is the angular frequency of the first laser. Similarly, the optical field 2 (depicted in red) couples the transition $|2\rangle \rightarrow |3\rangle$ with the Rabi frequency $\Omega_2$ and the detuning $\Delta_2=\omega_2-\omega_{32}$, where $\omega_{2}$ is the frequency of the second laser. $\Gamma_{31}$ and $\Gamma_{32}$ represent the population decay rates from $|3\rangle$ to $|1\rangle$ and $|2\rangle$, respectively. In addition, $\Gamma_{12}$ and $\Gamma_{21}$ denote incoherent population transfer rates between the two ground states $|1\rangle$ and $|2\rangle$, typically arising from collisions. 

In the Markovian limit, the atomic density matrix $\hat{\rho}$ evolves according to the following master equation
\begin{equation}~\label{eq: Liouvillian eq}
    \dot{\hat{\rho}} = -\frac{i}{\hbar} [\hat{H}, \hat{\rho}] + \mathcal{D}(\hat{\rho})\, .
\end{equation}
The system Hamiltonian $\hat{H}$ accounts for the coherent dynamics:
\begin{equation}
    \hat{H} = \Delta_1 \ket{1}\bra{1} + \delta \ket{2}\bra{2} + \left(\Omega_1 \ket{1}\bra{3} + \Omega_2 \ket{2}\bra{3} + \textrm{H.C.}\right),
\end{equation}
where $\delta = \Delta_1 - \Delta_2$, is the two-photon detuning. The dissipative term $\mathcal{D}(\hat{\rho})$ describes decoherence via jump operators $\hat{L}_m$ with rates $\Gamma_m$:
\begin{equation}~\label{eq: Lindblad}
    \mathcal{D}(\hat{\rho}) = \sum_m \frac{\Gamma_m}{2} \left(2\hat{L}_m \rho \hat{L}_m^\dagger - \{\hat{L}_m^\dagger \hat{L}_m , \hat{\rho}\}  \right)\, .
\end{equation}
For lifetime-broadened media, decoherence is dominated by spontaneous emission, modeled by $\hat{L}_1 = \sqrt{\Gamma_{31}} \ket{1}\bra{3}$ and $\hat{L}_2 = \sqrt{\Gamma_{32}} \ket{2}\bra{3}$. Collisional dephasing is included via $\hat{L}_3 = \sqrt{\Gamma_{12}} \ket{1}\bra{2}$. 

The steady-state elements $\rho_{mn}$ of the density matrix are obtained by solving Eq.~\eqref{eq: Liouvillian eq}:
\begin{align}
    0=\dot{\rho_{11}} &= - \frac{i}{2} (\Omega_1 \rho_{13} - \Omega_1^*\rho_{31}) -\Gamma_{12} \rho_{11} + \Gamma_{21} \rho_{22} + \Gamma_{31} \rho_{33} ~\label{eq: EOM1}\, ,\\
    0=\dot{\rho_{22}} &= - \frac{i}{2}(\Omega_2 \rho_{23}-\Omega_2^*\rho_{32}) + \Gamma_{12} \rho_{11} - \Gamma_{21} \rho_{22} + \Gamma_{32} \rho_{33} \,,\\
    0=\dot{\rho_{21}} &=-i(\Delta_2 - \Delta_1 - i \gamma_{12}) \rho_{21} -\frac{i}{2} (\Omega_1 \rho_{23} - \Omega_2^*\rho_{31}) \,, \\
    0=\dot{\rho_{31}} &=\frac{i}{2}\left[\Omega_1(\rho_{11} - \rho_{33})+ \Omega_2 \rho_{21}\right] + i(\Delta_1 + i \gamma_{13}) \rho_{31} , \\
    0=\dot{\rho_{32}} &= \frac{i}{2}\left[\Omega_2 (\rho_{22}-\rho_{33})+\Omega_1 \rho_{12} \right] + i(\Delta_2 + i \gamma_{23}) \rho_{32}. ~\label{eq: EOM5} 
\end{align}
where $\Gamma_3 = \Gamma_{31} + \Gamma_{32}$, and $\gamma_{12}$, $\gamma_{13}$, and $\gamma_{23}$ are the decoherence rates of non-diagonal elements. 
For cold atoms, $\gamma_{13} = (\Gamma_{3} + \Gamma_{12})/2$ and $\gamma_{23} =(\Gamma_{3} + \Gamma_{21})/2$. 
The ground-state dephasing rate can be expressed as $\gamma_{12} = \gamma_{12,0} + (\Gamma_{12} + \Gamma_{21})/2$, where $\gamma_{12,0}$ is the residual dephasing. 
Note that we only wrote the equations of motion for distinct density matrix entries, taking into account the conservation of particle numbers $\rho_{11} + \rho_{22} + \rho_{33} = 1$ and the hermiticity of the density matrix, i.e. $\rho_{mn} = \rho_{nm}^*$. 

Figure~\ref{fig:fig01}(b) shows the optical setup, where the two spatially overlapped beams propagate along the $z$ direction through a medium of length $L$. The beam propagation is governed by the slowly varying envelope equations:
\begin{equation}
    \begin{aligned}
        \frac{\partial}{\partial z} \Omega_n = i \frac{k_n}{\varepsilon_0\hbar} N |\mu_{n3}|^2 \rho_{3n}\, , 
    \end{aligned}~\label{eq:field1eq}
\end{equation}
where $n = 1,2$, $k_n = \omega_n/c$ is the wavenumber of the corresponding beam, $N$ is the atomic density, and $\mu_{n3}$ is the transition dipole moment between $|n\rangle$ and $|3\rangle$. 

Given input fields $\Omega_{1(2),\textrm{in}}$ at $z=0$, and after formally solving Eqs.~\eqref{eq: EOM1}-\eqref{eq:field1eq}, the output fields can be written as
\begin{equation}
    \begin{aligned}
        \Omega_{n,\textrm{out}} = \varphi_{n}( \Omega_{1,\textrm{in}},  \Omega_{2,\textrm{in}}) \, .
    \end{aligned}~\label{eq:field1}
\end{equation}
where $\varphi_{n}$ with $n=1,2$ denote nonlinear activation functions of $\Omega_{1,\textrm{in}}$ and $\Omega_{2,\textrm{in}}$. Figure~\ref{fig:fig01}(c) illustrates the simplified icon for this two-channel (two-input, two-output) nonlinear activation function unit.

%
\vspace{0.5cm}
\noindent\textbf{EIT in the weak-probe limit.}\label{subsec:EIT}
We begin with the simplest case of electromagnetically induced transparency (EIT) in a cold atomic ensemble, under the weak-probe approximation where $|\Omega_{1,\textrm{in}}| \ll |\Omega_{2,\textrm{in}}|$, and neglect collisional effects, i.e., $\Gamma_{12} = \Gamma_{21} = 0$. Under these conditions, the population remains predominantly in state $|1\rangle$, i.e., $\rho_{11} \approx 1$, and the non-depletion approximation applies. Since states $|2\rangle$ and $|3\rangle$ are barely populated, the control field remains unaffected: $\Omega_{2,\text{out}} = \Omega_{2,\text{in}} = \Omega_2$. Consequently, Eq.~\eqref{eq:field1eq} reduces to a linear propagation equation for the probe field, yielding
\begin{equation}
    \begin{aligned}
        \Omega_{1,\text{out}} = \Omega_{1,\text{in}} e^{i k_1 L \sqrt{1+\chi_1}} \cong \Omega_{1,\text{in}} e^{i k_1 L} e^{i k_1 \chi_1 L / 2},
    \end{aligned}~\label{eq:field1EIT}
\end{equation}
where the linear susceptibility $|\chi_1| \ll 1$ is given by
\begin{equation}
    \begin{aligned}
        \chi_1(\Delta_1) = \frac{4N|\mu_{13}|^2 (\Delta_1 - \Delta_2 + i\gamma_{12}) / (\varepsilon_0 \hbar)}{|\Omega_2|^2 - 4 (\Delta_1 + i\gamma_{13})(\Delta_1 - \Delta_2 + i\gamma_{12})}.
    \end{aligned}~\label{eq:chiEIT}
\end{equation}
In the ideal case with no ground-state dephasing ($\gamma_{12} = 0$), the susceptibility vanishes at two-photon resonance ($\Delta_1 = \Delta_2$), yielding $\chi_1(0) = 0$, and thus full transparency for the probe field 1. Notably, this transparency is independent of the strength of the control field, as even an arbitrarily weak control field can render the medium transparent for a much weaker probe field. To enable nonlinear activation, we consider a finite dephasing rate $\gamma_{12} \neq 0$, which introduces an imaginary component to the susceptibility:
\begin{equation}
    \begin{aligned}
        \chi_1(0) = \frac{i4N|\mu_{13}|^2 \gamma_{12} / (\varepsilon_0 \hbar)}{|\Omega_2|^2 + 4\gamma_{13} \gamma_{12}}\, .
    \end{aligned}~\label{eq:chiEIT0}
\end{equation}
Substituting Eq.~\eqref{eq:chiEIT0} into Eq.~\eqref{eq:field1EIT}, the output field becomes
\begin{equation}
    \begin{aligned}
        \Omega_{1,\text{out}}(\Delta_1 = 0) = \Omega_{1,\text{in}} e^{i k_1 L} e^{- \textrm{OD} \frac{2\gamma_{12} \gamma_{13}}{|\Omega_2|^2 + 4\gamma_{13} \gamma_{12}}},
    \end{aligned}~\label{eq:field1EIT1}
\end{equation}
or equivalently
\begin{equation}
    \begin{aligned}
        |\Omega_{1,\text{out}}(\Delta_1 = 0)|^2 = |\Omega_{1,\text{in}}|^2 e^{- \textrm{OD} \frac{4\gamma_{12} \gamma_{13}}{|\Omega_2|^2 + 4\gamma_{13} \gamma_{12}}},
    \end{aligned}~\label{eq:EIT1}
\end{equation}
where OD = $N\sigma_{13}L$ is the optical depth for the $|1\rangle \rightarrow |3\rangle$ transition without the control field, and $\sigma_{13} = k_1 |\mu_{13}|^2 / (\varepsilon_0 \hbar \gamma_{13})$ is the resonant absorption cross section. Equation~\eqref{eq:EIT1} shows that the probe output power is nonlinearly controlled by the control field intensity $|\Omega_2|^2$, forming the basis of an optical activation function with $\Omega_{2,\text{in}}$ as input and $\Omega_{1,\text{out}}$ as output, as previously demonstrated~\cite{zuo2019all, PRApplied-Du-2021}. Here, we call field 1 the probe field and field 2 the control field.

\begin{figure}[t]
    \centering
    \includegraphics[width=\linewidth]{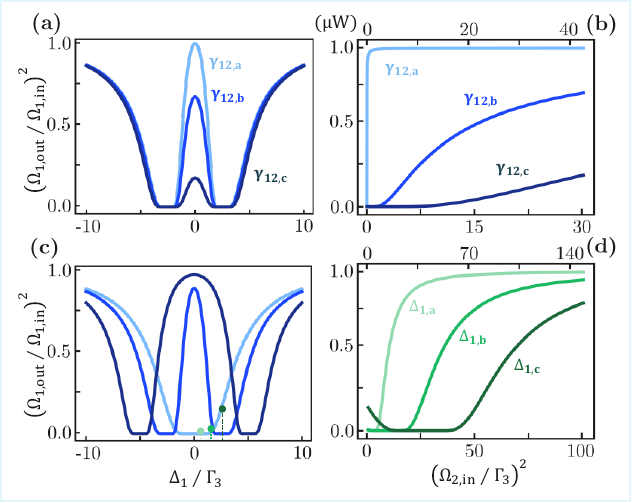}
    \caption{\textbf{EIT in the weak probe regime.} $\Delta_2=0$. 
    (a) EIT transmission spectra for varying ground-state decoherence rates, with $(\gamma_{12,a}, \gamma_{12,b}, \gamma_{12,c}) = (0.0001, 0.1, 0.5)\,\Gamma_3$, and a fixed control field $\Omega_2 = 5\,\Gamma_3$. 
    (b) EIT nonlinear activation functions with different ground-state decoherence rates $\gamma_{12}$, expressed as the cross-nonlinearity between the probe output $|\Omega_{1,out}|^2$ and the control input $|\Omega_{2}|^2$ [Eq.~(\ref{eq:EIT1})].
    (c) EIT transmission spectra for different control Rabi frequencies $\Omega_2 = (0, 5, 10)\,\Gamma_3$ at fixed $\gamma_{12} = 0.03\,\Gamma_3$. Green markers denote selected probe detunings $(\Delta_{1,a}, \Delta_{1,b}, \Delta_{1,c}) = (0.5, 1.5, 2.5)\,\Gamma_3$. 
    (d) Corresponding nonlinear activation functions for the three detuning values shown in (c), illustrating tunable sigmoid-like response.
    }
    \label{fig:fig02}
\end{figure}

To illustrate this nonlinear mechanism, we simulate transmission through a cold $^{87}$Rb atomic ensemble, as in Refs.~\cite{zuo2019all, PRApplied-Du-2021}. We use states $|1\rangle = |5S_{1/2}, F=1\rangle$, $|2\rangle = |5S_{1/2}, F=2\rangle$, and $|3\rangle = |5P_{1/2}, F=2\rangle$, setting $\Delta_2 = 0$, $\Gamma_3 = 2\pi \times 6$ MHz, and $\gamma_{13} = \gamma_{23} = \Gamma_3 / 2$. To estimate the required laser powers, we assume the medium length $L=2$ cm. We take a beam radius of 71 $\mu$m such that the diffraction length, for the wavelength of the $^{87}$Rb D1 line of 795 nm, is $b=4$ cm, which is much longer than the medium length. See Appendix \ref{Appendix: Numerical Simulation} for more details about the numerical simulation.

Figure~\ref{fig:fig02}(a) shows the probe transmission spectrum for OD = 50 and three dephasing rates: $\gamma_{12,a} = 0.0001 \Gamma_3$, $\gamma_{12,b} = 0.1 \Gamma_3$, and $\gamma_{12,c} = 0.5 \Gamma_3$. At resonance ($\Delta_1 = 0$), the transmission increases with decreasing $\gamma_{12}$ and increasing control field power. As shown in Fig.~\ref{fig:fig02}(b), the EIT peak saturates at unity for large $|\Omega_2|^2$. The slope of the transmission curve depends on $\gamma_{12}$, consistent with Eq.~\eqref{eq:EIT1} and prior experiments~\cite{zuo2019all, PRApplied-Du-2021}.

To expand beyond the resonance EIT explored in prior work, we demonstrate that the turning point of the nonlinear activation function can be tuned by the probe detuning $\Delta_1$. As shown in Fig.~\ref{fig:fig02}(c), for $\Delta_1 \neq 0$, the probe initially experiences absorption. As the control field increases, the EIT bandwidth broadens and eventually encompasses the probe detuning, resulting in transparency. Figure~\ref{fig:fig02}(d) shows the resulting sigmoid-like activation functions with tunable thresholds, enabling flexible reconfiguration—a capability not previously demonstrated experimentally.

\begin{figure}[t]
    \centering
    \includegraphics[width=\linewidth]{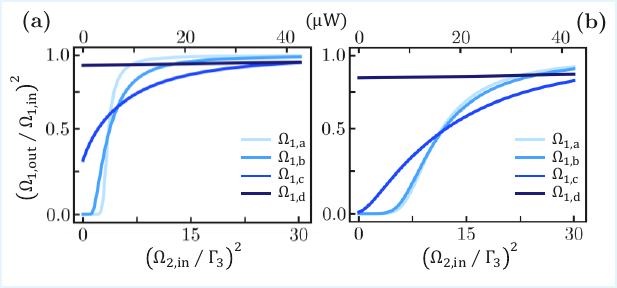}
    \caption{\textbf{Power effect on the resonant probe transmission ($\Delta_1=\Delta_2=0$).} Control-probe nonlinear functions with different probe inputs $(\Omega_{1,a}, \Omega_{1,b}, \Omega_{1,c}, \Omega_{1,d})=(10^{-5}, 1, 3, 10) \Gamma_3$ for (a) $\Gamma_{12}=0.1\Gamma_3$ and (b) $\Gamma_{12}=\Gamma_3$.}
    \label{fig:fig03}
\end{figure}


%
\begin{figure}[t]
    \centering
    \includegraphics[width=\linewidth]{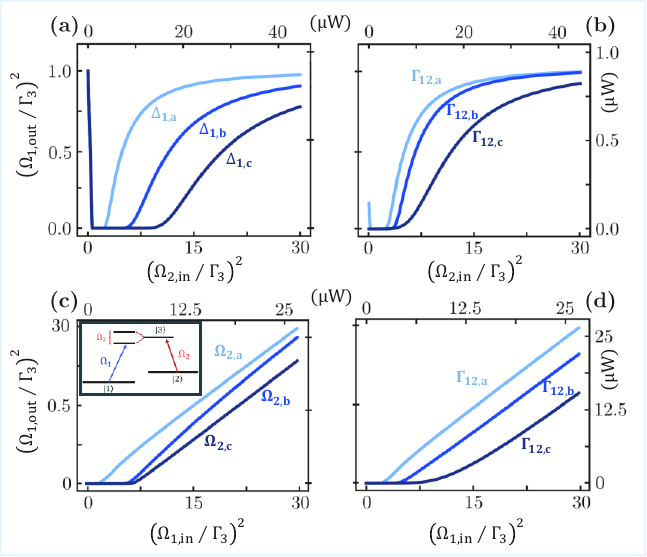}
\caption{\textbf{Nonlinear optical activation functions with comparable driving fields.} $\Delta_2 = 0$. (a) Control-probe cross nonlinear activation functions between $|\Omega_{1,out}|^2$ and $|\Omega_{2,in}|^2$ with different probe detuning: $\Delta_1=(1/3, ~ 2/3, ~1) \Gamma_3$. Other parameters: $\Omega_{1,\mathrm{in}}=\Gamma_3$, and $\Gamma_{12}=0$. (b) Control-probe cross nonlinear activation functions between $|\Omega_{1,\mathrm{out}}|^2$ and $|\Omega_{2,\mathrm{in}}|^2$ with different ground state population transfer rate: $\Gamma_{12} = (0.01,~ 0.1,~ 1) \Gamma_3$. Other parameters: $\Omega_1 = \Gamma_3$, and $\Delta_1 = 1/3 \Gamma_3$. (c) Probe self nonlinear activation functions with different control powers: $(\Omega_{2,a},\Omega_{2,b},\Omega_{2,b})=(1,~3, ~10) \Gamma_3$. Other parameters: $\Delta_1 = \Omega_2/2$, and $\Gamma_{12}=0$. (d) Probe self nonlinear activation functions with different ground state population transfer rate: $(\Gamma_{12,a},\Gamma_{12,b},\Gamma_{12,c}) = (0.01,~ 0.1,~ 1) \Gamma_3$. Other parameters are $\Omega_2=\Gamma_3$, and $\Delta_1=\Omega_2/2$.} 
    \label{fig:fig04}
\end{figure}
\begin{figure}[t]
    \centering
    \includegraphics[width=\linewidth]{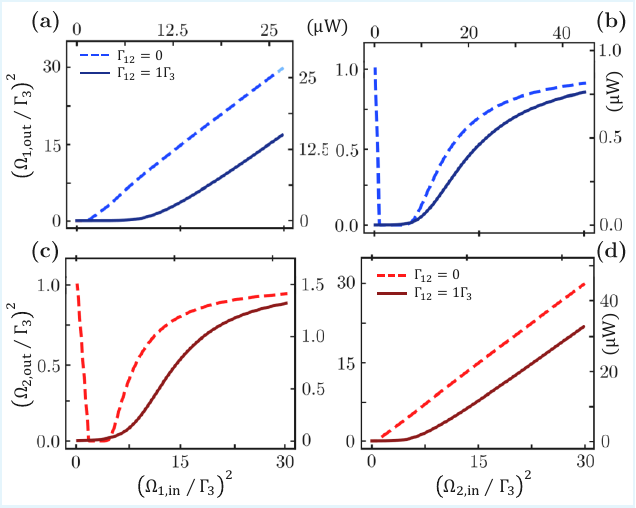}
    \caption{\textbf{ Two-channel (2-input $\times$ 2-output) nonlinear activation functions with self- and cross-nonlinearity in a lifetime-broadened atomic medium.} (a) Self- and (b) cross-nonlinearity of input 1, respectively. (c) Cross- and (d) self-nonlinearity of input 2, respectively. We set $\Omega_{1,\mathrm{in}}=\Gamma_3$ in (b) and $\Omega_{2,\mathrm{in}}=\Gamma_3$ in (c). $\Delta_1 = - \Delta_2 = 1/3\Gamma_3$.}
    \label{fig:fig05}
\end{figure}

\vspace{0.5cm}
\noindent\textbf{Two-channel nonlinear optical activation functions.}~\label{subsec:engineering} In the above EIT-based nonlinear activation function described above, the strong control field (field 2) serves as the input, and the weak probe field (field 1) is the output, with atomic population primarily in state $|1\rangle$. In a multilayer AONN architecture, however, the roles of control and probe fields must alternate between layers. This creates a challenge: the output probe field from one layer becomes the control field for the next, while a new, even weaker probe field must be introduced. Due to the EIT condition $|\Omega_p| \ll |\Omega_c|$, the signal power diminishes with increasing network depth. In an ideal EIT system with zero ground-state dephasing ($\gamma_{12}=0$), such a layered scheme remains functional as long as the weak-probe condition holds. In practice, however, a finite ground-state dephasing rate $\gamma_{12} \neq 0$ imposes a power threshold on the control field, requiring $|\Omega_c|^2 > 4~\mathrm{OD}~\gamma_{12} \gamma_{13}$ to maintain transparency. This limits the maximum achievable depth of the AONN.

To overcome this limitation, we propose a two-field scheme in which both laser fields have comparable intensities. Although no analytical expression exists for this case, a qualitative understanding can be drawn. When $\Omega_1 \approx \Omega_2$, the atomic population distribution between states $|1\rangle$ and $|2\rangle$ becomes dependent on the relative field strengths. As $\Omega_2$ is increased from zero and remains weaker than $\Omega_1$, atoms are increasingly pumped into state $|1\rangle$, enhancing the absorption of field 1. We refer to this process as \emph{electromagnetically induced absorption} (EIA). As $\Omega_2$ exceeds $\Omega_1$, EIT becomes dominant, and absorption of field 1 decreases. This transition from EIA to EIT leads to a sigmoid-like nonlinear activation function, with the turning point determined by the crossover. Figure~\ref{fig:fig03} shows simulated resonant ($\Delta_1 = \Delta_2 = 0$) probe transmissions $(\Omega_{1,\mathrm{out}}/\Omega_{1,\mathrm{in}})^2$ for various input probe intensities. Panel (a) corresponds to a small ground-state population transfer rate $\Gamma_{12} = 0.1 \Gamma_3$, while panel (b) uses $\Gamma_{12} = \Gamma_3$. In both cases, the sigmoid nonlinearity breaks down for $\Omega_{1,\mathrm{in}} > 3\Gamma_3$ but remains effective at moderate intensities ($\Omega_{1,\mathrm{in}} \leq 3\Gamma_3$), comparable to $\Omega_2$. The presence of finite $\Gamma_{12}$ enhances absorption at lower coupling powers, effectively mitigating the degradation in sigmoid behavior due to increasing probe intensity. However, above a certain power threshold, the nonlinear response becomes ineffective.

We next examine the self-nonlinearity of field 1 in the presence of a fixed field 2. Initially, when $|\Omega_1| < |\Omega_2|$, the atomic population resides primarily in state $|1\rangle$, resulting in high absorption. As $\Omega_1$ increases, atoms are pumped out of state $|1\rangle$, reducing absorption and giving rise to a ReLU-like activation function. Unlike simple saturation in a two-level system, this three-level quantum interference allows for tunable nonlinear responses, as verified in numerical simulations.

Figure~\ref{fig:fig04} presents numerical results for nonlinear activation functions with comparable fields. Panel (a) shows sigmoid-type responses—output probe power $|\Omega_{1,\mathrm{out}}|^2$ as a function of control field power $|\Omega_{2,\mathrm{in}}|^2$—for different probe detunings $\Delta_1$, under $\Gamma_{12} = \Gamma_{21} = 0$ and $\Delta_2 = 0$. Similar to the EIT case in Fig.~\ref{fig:fig02}(d), the turning point of the nonlinearity shifts to higher control powers as $\Delta_1$ increases. A narrow transmission spike near $\Omega_{2,\mathrm{in}} = 0$ is caused by the pumping effect of the strong field 1, and is suppressed when finite $\Gamma_{12}$ is included, as shown in panel (b). Panel (c) shows ReLU-type self-nonlinear activation functions for field 1, under varying control powers. The turning point of the response shifts with $|\Omega_{2,\mathrm{in}}|^2$, offering dynamic control. When field 2 is on resonance ($\Delta_2 = 0$), it induces a dressed-state splitting of level $|3\rangle$ into two levels separated by $\hbar\Omega_2$. The optimal detuning for field 1 is then $\Delta_1 = \Omega_2/2$ to maximize interaction. Panel (d) illustrates how different values of $\Gamma_{12}$ affect the response at this detuning.

With both fields at comparable strengths, we further demonstrate two-channel (2-input × 2-output) nonlinear activation functions, suitable for multi-input, multi-output (MIMO) photonic architectures. Figure~\ref{fig:fig05} displays the output intensities of each beam as functions of their own and the other beam’s input intensities under symmetric detunings $\Delta_1 = -\Delta_2 = \Gamma_3/3$. The self-nonlinear responses in panels (a) and (d) are similar for both fields. The cross-nonlinearities in panels (b) and (c) confirm that either beam can modulate the other, enabling flexible design of reconfigurable, dual-channel activation units.

\section{Doppler-Broadened Medium}\label{sec:doppler-broadened}
%
The preceding results highlight the versatility and tunability of nonlinear activation functions in lifetime-broadened media. While our simulations were based on cold atoms, the underlying principles are broadly applicable to other atom-like platforms, such as quantum dots and solid-state defect centers~\cite{michler2003single, aharonovich2016solid}. In parallel, thermal atomic vapors—despite exhibiting Doppler broadening, elevated dephasing, and collisional effects—remain attractive candidates for scalable implementations due to their simplicity and integrability. Recent advances in coupling thermal vapors with nanophotonic devices open pathways for hybrid architectures that combine nonlinear activation with on-chip linear photonic circuits~\cite{Alaeian2020, Skljarow2020, Skljarow2022}. To evaluate the robustness of the aforementioned results to decoherence and Doppler broadening, we investigate the achievable nonlinearity with a thermal vapor platform. 

\begin{figure*}[t]
    \centering
    \includegraphics[width=0.85\linewidth]{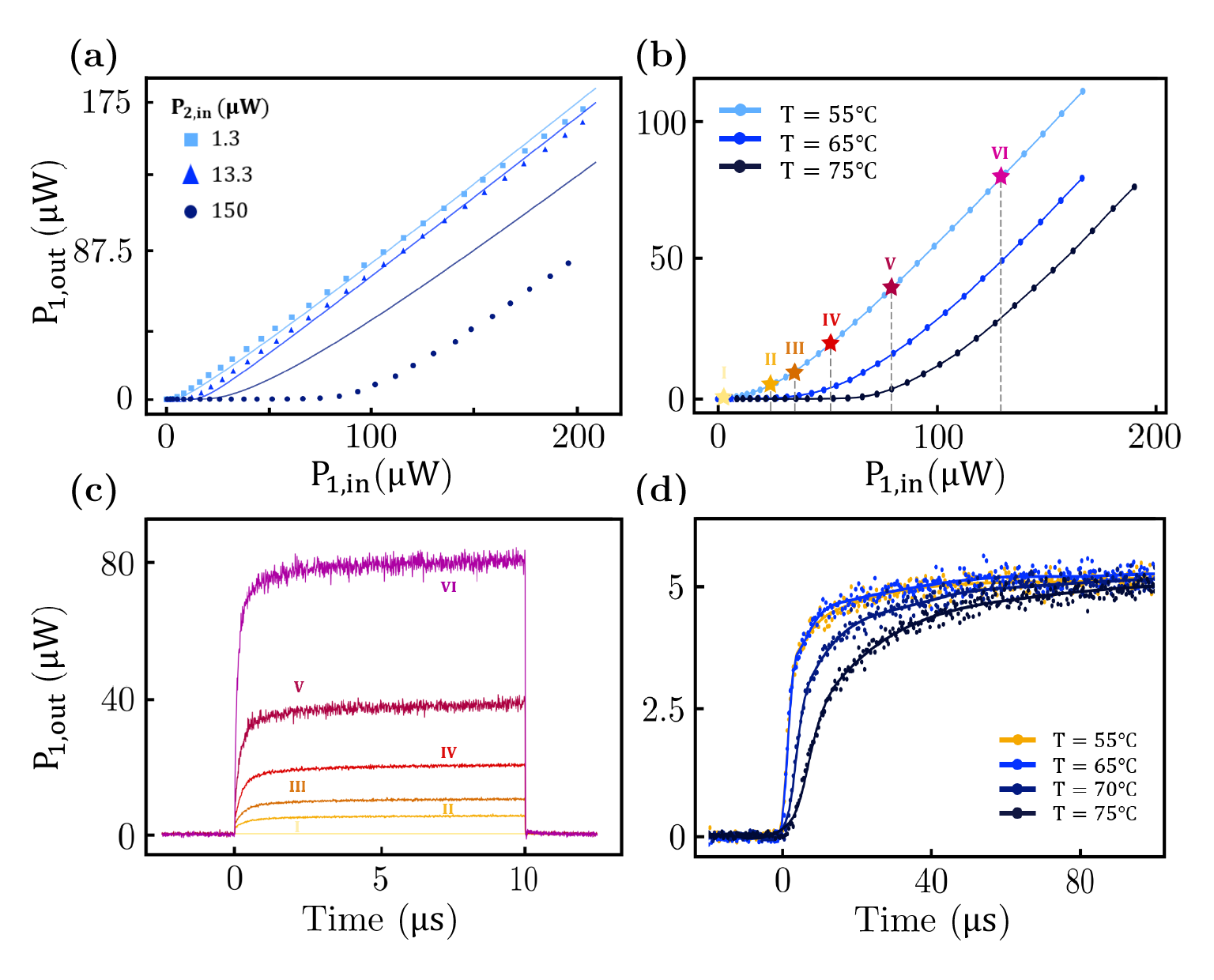}
    \caption{\textbf{ReLU nonlinear activation functions for channel 1 with a Rb vapor cell.}  All measurements were performed with detunings set to $\Delta_1 = \Omega_{2,\mathrm{in}}/2$ and $\Delta_2 = 0$. 
(a) Measured ReLU activation functions (dots) for varying control beam powers at fixed temperature $T = 75^\circ$C. Solid lines represent numerical simulations fitted with $\Gamma_{12} = 2\pi \times (0.09,\, 0.132,\, 0.6)$~MHz. 
(b) Temperature-dependent ReLU curves recorded at a fixed control power of 150~$\mu$W. Asterisks I–VI mark input powers used in transient measurements. 
(c) Transient response of the system under 10~$\mu$s rectangular probe pulses corresponding to the power levels marked in (b). Control power is fixed at 150~$\mu$W. 
(d) Measured rise times at different cell temperatures, recorded at the ReLU turning point using 400~$\mu$s pulses to ensure saturation. 
    } 
    \label{fig:fig06}
\end{figure*}

With the same configuration shown in Fig.~\ref{fig:fig01}(a),(b), along the beams, atoms follow the one-dimensional Maxwell-Boltzmann velocity distribution
\begin{equation}
    f(v) = \frac{1}{v_\textrm{th} \sqrt{\pi}} e^{-\left(v/v_\textrm{th}\right)^2}\, ,
\end{equation}
where $v_\textrm{th}$ is the most probable thermal speed related to the atomic mass $m$ and temperature $T$ as 
\begin{equation}
    v_\textrm{th} = \sqrt{2 k_B T/m}\, ,
\end{equation}
with $k_B$ the Boltzmann constant. An atom moving at velocity $v$ experiences a Doppler shift, such that a laser field with nominal detuning $\Delta_0$ acquires an effective detuning:
\begin{equation}~\label{eq: Doppler}
    \Delta_n(v) = \Delta_0 - k_nv\, ,
\end{equation}
for $k_n=\omega_n/c$ being the wave number of field $n$. The averaged $\bar{\rho}_{mn}$ can then be written as the weighted integral of the density matrices of all velocity classes as 
\begin{equation}
    \bar{\rho}_{mn} = \int_{-\infty}^{+ \infty} dv~ \rho_\textrm{mn}(\Omega , \Delta_n(v)) f(v)\, ,
\end{equation}
where $\rho_\textrm{mn}(v)$ is the steady-state solution of the density matrix determined through Eqs.~\eqref{eq: EOM1}-\eqref{eq: EOM5}. 

To experimentally verify our theoretical predictions, we use a cylindrical glass cell filled with pure $^{87}$Rb vapor, with a length of 2 cm and a diameter of 2 cm. Both laser beams are focused into the cell with Gaussian waist radius (at $1/e^2$ intensity) measured to be 75~$\mu$m. Additional experimental details are provided in Appendix \ref{Appendix: Experimental Details}.

\vspace{0.5cm}
\noindent\textbf{ReLU nonlinear activations and transient responses. }\label{subsec:RELU transient}
Figure~\ref{fig:fig06}(a) shows the measured ReLU-type nonlinear activation function, where the output power P$_{1,\mathrm{out}}$ of beam 1 is plotted against its input power P$_{1,\mathrm{in}}$ at $T = 75^\circ$C for three different input powers of beam 2: P$_{2,\mathrm{in},a} = 1.3~\mu$W (blue squares), P$_{2,\mathrm{in},b} = 13.3~\mu$W (blue triangles), and P$_{2,\mathrm{in},c} = 150~\mu$W (blue circles). As expected, increasing the control power shifts the turning point of the ReLU function toward higher P$_{1,\mathrm{in}}$. Solid curves represent numerical simulations fitted with ground-state population transfer rates $\Gamma_{12} = 2\pi \times (0.09, 0.132, 0.6)$~MHz, respectively. Notably, a single constant $\Gamma_{12}$ cannot reproduce all experimental curves simultaneously. In particular, the highest-power case (P$_{2,\mathrm{in},c}$) deviates from the model, likely due to the simplified three-level approximation neglecting hyperfine and Zeeman substructure. Nonetheless, the experimental data show good qualitative agreement with theoretical predictions.

The nonlinear activation response is also tunable via temperature, as shown in Fig.~\ref{fig:fig06}(b), where ReLU curves are measured at $T = 55^\circ$C, $65^\circ$C, and $75^\circ$C. As temperature increases, the turning point shifts to higher power, consistent with the scaling of saturation power with optical density. At lower temperatures, activation thresholds can be as low as $10~\mu$W.

In addition to the tunability, the system's temporal response is critical for evaluating its computational performance. Figure~\ref{fig:fig06}(c) shows the transient dynamics of the output signal for laser 1 at various input powers, corresponding to selected points in Fig.~\ref{fig:fig06}(b). The characteristic response time is approximately 1~$\mu$s, with faster transitions observed at higher powers due to increased optical pumping rates. We find that temperature also plays a key role in modulating response speed. Figure~\ref{fig:fig06}(d) compares the transient response times at $T = 55^\circ$C, $65^\circ$C, $70^\circ$C, and $75^\circ$C, with $P_{1,\mathrm{out}} = 5~\mu$W. While the response time remains nearly constant between $55^\circ$C and $65^\circ$C, a marked change occurs between $65^\circ$C and $75^\circ$C. This temperature dependence, which contrasts with the stronger temperature sensitivity of the ReLU turning point itself, provides additional control for optimizing device performance. Depending on the specific requirements of various applications, optimal settings for temperature and power can be selectively determined. 

The system’s intrinsic response time imposes a fundamental limit on the operating frequency of AONNs based on atomic media. While classical electronic processors operate in the gigahertz clock speed -- A general computing process normally takes many clock cycles, our system operates on the order of megahertz—several orders of magnitude slower in clock speed. However, the overall optical response time of an AONN scales with only the square root of the network depth but independent of its transverse size. Thus, for sufficiently large-scale networks, the parallelism inherent to AONNs can enable overall processing speeds that outperform electronic counterparts. Moreover, we find that the falling response time of a rectangular pulse is as short as 1$\sim$2 ns (limited by the detector resolution) as shown in Fig. ~\ref{fig:fig06}(c), which is determined by the bandwidth of Doppler broadening. This finding suggests that we can operate the optical neurons in the transient response mode (such as about 100 MHz repetition rate) for high speed computing instead of steady-state nonlinearity. As such, the observed saturation and rise times do not impose a practical limitation on the scalability or applicability of the proposed scheme. The vapor cell used in this study contains pure $^{87}$Rb. Introducing specific buffer gases may help reduce the response time \cite{Krishnamurthy:15}, although this effect warrants further investigation.

\begin{figure}[h!]
    \centering
    \includegraphics[width= 0.9\linewidth]{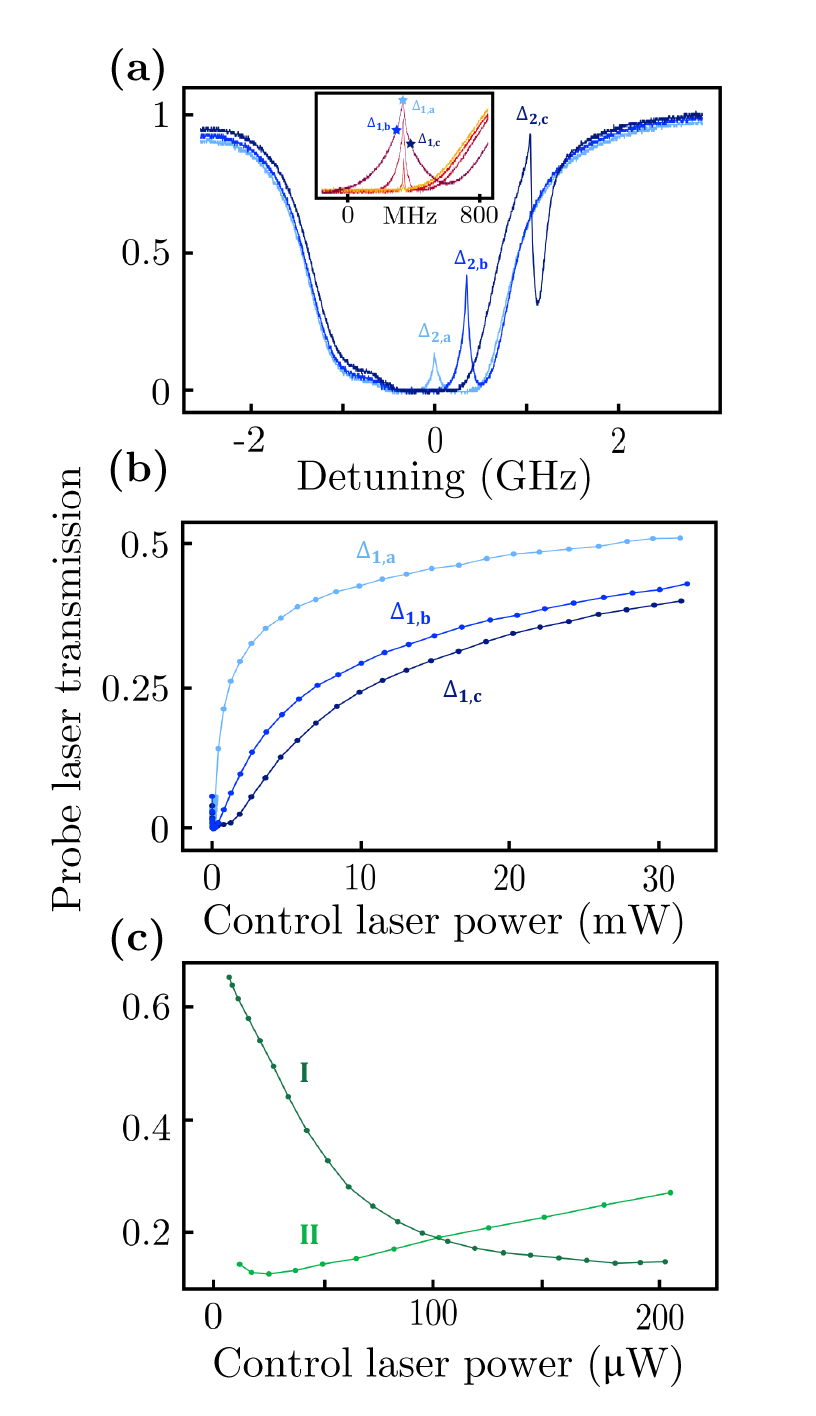}
    \caption{\textbf{Sigmoid-like nonlinear activation functions via cross-nonlinearity in a rubidium vapor cell.} 
(a) EIT transmission spectra for probe laser 1 as a function of detuning $\Delta_1$, with fixed input power $P_{1,\mathrm{in}} = 1.3~\mu$W. Control field detunings are set to $\Delta_{2,a} = 0$ MHz, $\Delta_{2,b} = 2\pi \times 352$ MHz, and $\Delta_{2,c} = 2\pi \times 1024$ MHz. The inset shows EIT responses at control powers $P_{2,\mathrm{in}} = 0.1$, 0.5, 5, and 20 mW (yellow to crimson). Asterisks mark the detunings used in panel (b). 
(b) Sigmoid activation functions measured at $\Delta_{1,a} = 0$ MHz, $\Delta_{1,b} = 2\pi \times (-20)$ MHz, and $\Delta_{1,c} = 2\pi \times 20$ MHz, relative to a fixed control detuning $\Delta_{2,b} = 2\pi \times 352$ MHz. 
(c) Nonlinear transmission behavior in the low control power regime. Curve I shows self-nonlinearity (ReLU-type) of laser 1 at $T = 75^\circ$C. Curve II illustrates cross-nonlinearity dominated by EIA at low optical depth, followed by a transition to EIT at higher control powers. This transition enables low-power switching between ReLU- and sigmoid-like responses.}
    \label{fig:fig07}
\end{figure}

\vspace{0.5cm}
\noindent\textbf{Sigmoid-like nonlinear activation functions. }\label{subsec:Sigmoid Doppler}
Figure~\ref{fig:fig07} presents the nonlinear activation behavior arising from EIT-based cross-nonlinearity. In panel (a), we show EIT transmission spectra under various control field detunings $\Delta_{2,a,b,c}$. The corresponding nonlinear activation functions are shown in panel (b), where the probe transmission is plotted versus control power for three probe detunings $\Delta_{1,a,b,c}$. At small detuning $\Delta_{1,a}$, saturation is reached at approximately 10~mW. For larger detunings $\Delta_{1,b}$ and $\Delta_{1,c}$, higher coupling powers in the 20–30~mW range are required to achieve similar levels of transparency. While the resulting sigmoid-like curves resemble those observed in lifetime-limited media [Fig.~\ref{fig:fig02}(d)], Doppler broadening alters key characteristics such as steepness, threshold location, and peak transmission. These differences highlight the distinct activation dynamics of a thermal atomic ensemble.

Beyond conventional sigmoid and ReLU functions, additional nonlinear profiles can be engineered, as illustrated in Fig.~\ref{fig:fig07}(c). At elevated temperatures, where EIA dominates for low input power $P_{2,\mathrm{in}}$, the system exhibits a reversed ReLU response (e.g., curve I). As input power increases, the activation behavior transitions to that shown in Fig.~\ref{fig:fig07}(b). At lower temperatures, we observe a clear evolution from EIA to EIT with increasing control power. This transition offers an alternative route to implement ReLU-type activation within a low-power regime ($\leq 200~\mu$W) and enables convenient switching between sigmoid and ReLU nonlinearities. While our AONN implementation focuses on conventional activation functions, these EIA-to-EIT transitions may enable additional reconfigurable architectures and merit further investigation for advanced photonic neural network designs.

\begin{figure}[h!]
    \centering
    \includegraphics[width=0.93\linewidth]{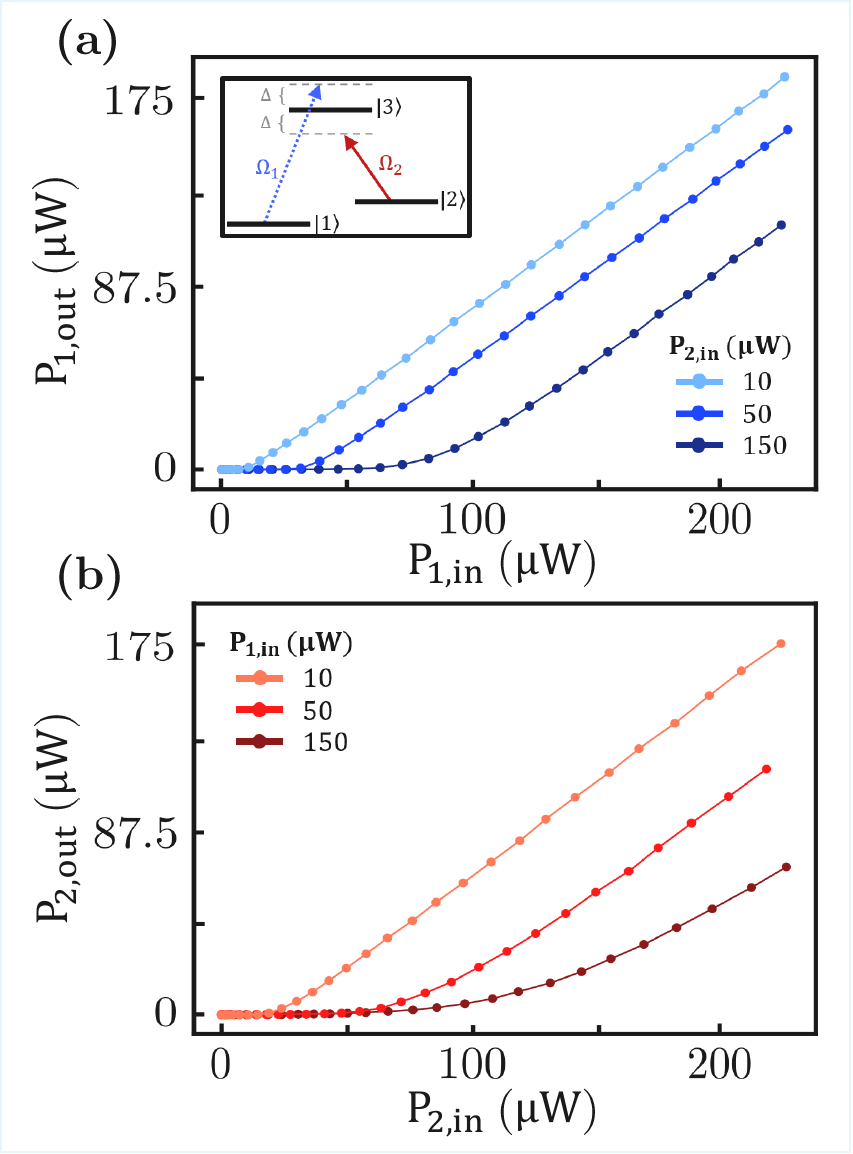}
    \caption{\textbf{Symmetric two-channel nonlinear activation functions in a rubidium vapor cell.} 
Measurements are performed with detunings $\Delta_1 = -\Delta_2 = 2\pi \times 2$~MHz. 
(a) ReLU-type activation behavior in channel 1 for varying input powers in channel 2 (P$_{2,\mathrm{in}}$). 
(b) ReLU-type activation behavior in channel 2 for varying input powers in channel 1 (P$_{1,\mathrm{in}}$).}
    \label{fig:fig08}
\end{figure}

\vspace{0.5cm}
\noindent\textbf{Two-channel nonlinear activations. }\label{subsec:Two-channel Doppler}
Using two optical fields with comparable powers not only improves energy efficiency but also enables the realization of symmetric dual-channel (2-input × 2-output) nonlinear activation functions. Experimental validation of this concept is shown in Fig.~\ref{fig:fig08}. To achieve symmetric performance between the two channels, we set the detunings to $\Delta_1 = -\Delta_2 = 2\pi \times 2$~MHz. Under these conditions, both channels exhibit ReLU-like activation behavior in response to variations in the other channel’s input, without the need to change laser frequencies. This symmetric, tunable nonlinearity offers new opportunities for multi-channel AONN architectures, as discussed in the Outlook section. Importantly, because the ReLU response here is mediated by population transfer rather than EIT, the behavior is robust against Doppler broadening and closely resembles that of cold atom systems shown in Fig.~\ref{fig:fig05}.

\begin{figure}[h!]
    \centering
    \includegraphics[width=0.95\linewidth]{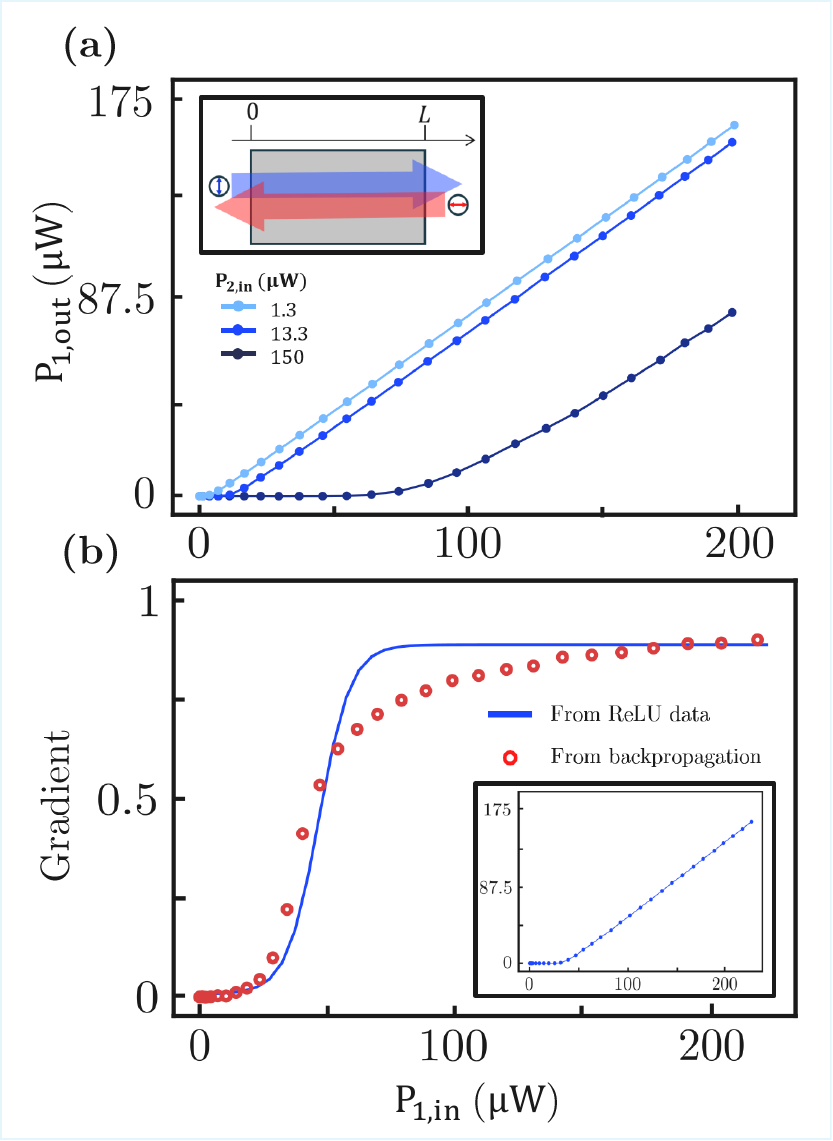}
    \caption{\textbf{Nonlinear activation and gradient measurement with counter-propagating beams.} 
(a) ReLU activation functions measured under a counter-propagating geometry, using the same parameters as Fig.~\ref{fig:fig06}(a). Inset: schematic of the counter-propagating setup with orthogonal linear polarizations. 
(b) Gradient extracted from a backward-propagating weak probe input (P$_{\mathrm{probe}} = 5~\mu$W, red circles) agrees well with the derivative of the measured ReLU function (inset), demonstrating direct optical gradient readout for backpropagation.}
    \label{fig:fig09}
\end{figure}

\vspace{0.5cm}
\noindent\textbf{Counter-propagation and optical gradient measurement. }\label{subsec:Counter-propagation Doppler}
Although our theoretical framework assumes two co-propagating and spatially overlapped laser beams, we experimentally confirm that ReLU-type nonlinear activation functions are also preserved under a counter-propagating configuration. As shown in Fig.~\ref{fig:fig09}(a), the measured activation curve closely matches the results from the co-propagating geometry presented in Fig.~\ref{fig:fig06}(a). The ability to implement both co-propagating and counter-propagating geometries provides flexibility in optical system design and is particularly advantageous for realizing optical backpropagation training in AONNs~\cite{10.1117/1.AP.7.1.016004, Guo:21, Hughes:18}. To demonstrate this capability, we introduce a weak (5~$\mu$W) backward-propagating probe beam in the co-propagating setup. Its transmitted signal directly measures the local gradient of the activation function. As shown in Fig.~\ref{fig:fig09}(b), the experimentally measured gradient (red circles) agrees well with the derivative extracted from the ReLU curve, confirming the feasibility of optical gradient computation. Minor discrepancies at higher powers may arise from imperfect beam alignment or polarization mismatch.

\begin{figure}[h!]
    \centering
    \includegraphics[width=0.95\linewidth]{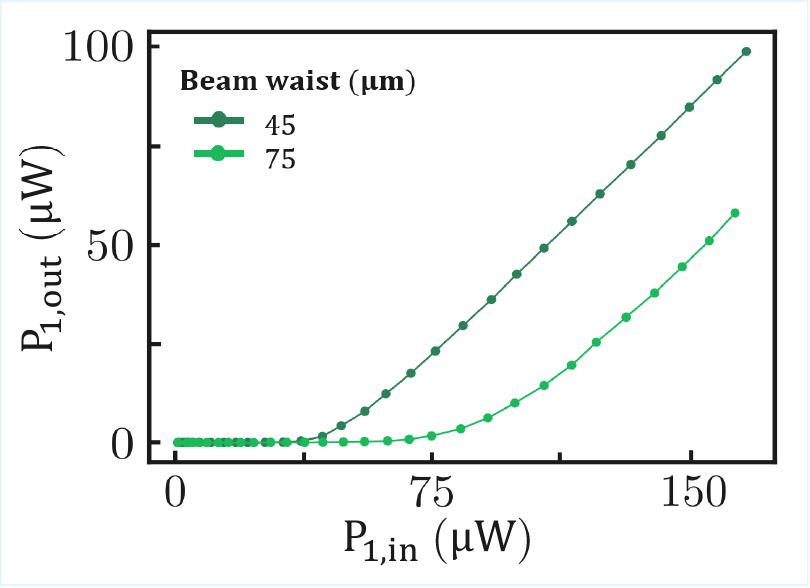}
   \caption{\textbf{ReLU nonlinear activation functions for different beam waists.} 
Comparison of activation curves for beam waist radii of 75~$\mu$m and 45~$\mu$m. Detunings are set to $\Delta_1 = -\Delta_2 = 2\pi \times 2$~MHz. In both cases, input powers of laser beam 2 were adjusted to maintain a constant Rabi frequency $\Omega_{2,in}$. Smaller beam size yields a lower activation threshold due to enhanced field intensity.}
    \label{fig:fig10}
\end{figure}

\vspace{0.5cm}
\noindent\textbf{Beam size and power reduction. }\label{subsec:Beam sizze Doppler}
In the preceding experiments, the laser beam waist was set to 75~$\mu$m to approximate a plane-wave configuration, thereby aligning with our theoretical model. However, further power reduction is possible by decreasing the beam size, even if the plane-wave assumption no longer strictly holds. Figure~\ref{fig:fig10} compares ReLU-type nonlinear activation functions measured using beam waists of 75~$\mu$m and 45~$\mu$m. In both cases, input powers of laser beam 2 were adjusted to maintain a constant Rabi frequency $\Omega_{2,in}$. As shown in the figure, the activation threshold shifts significantly: the turning point is reduced from approximately 80~$\mu$W to 40~$\mu$W with the smaller beam waist. This demonstrates that smaller beams can substantially lower the power required per optical neuron. Although a reduced beam size increases diffraction, this can be compensated by shortening the cell length and increasing cell temperature to preserve sufficient optical depth. This trade-off offers a practical path toward implementing ultra-low-power AONNs with compact physical footprints.

\section{Discussion and Outlook}\label{sec:conclusion}

\begin{figure}[t]
    \centering
    \includegraphics[width=\linewidth]{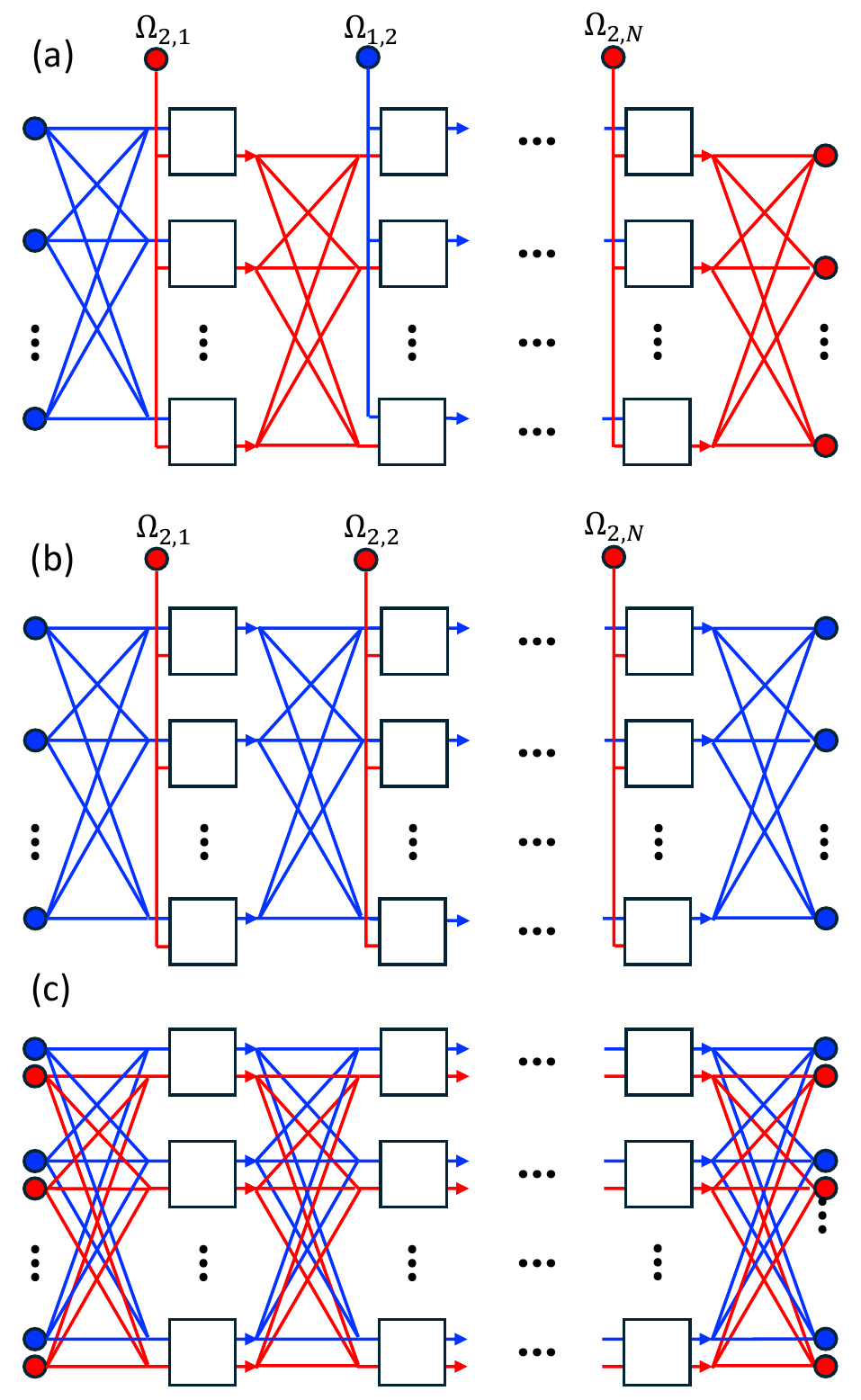}
    \caption{\textbf{Architectures for multilayer deep AONNs.} 
    (a) AONN based on single-channel control–probe sigmoid nonlinear activation functions. 
    (b) AONN using single-channel ReLU-type self-nonlinear activation functions. 
    (c) MIMO AONN architecture enabled by two-channel (2-input × 2-output) nonlinear activation functions.}
    \label{fig:fig11}
\end{figure}
%

In this work, we demonstrate the engineering of low-power optical nonlinear activation functions using quantum interference in three-level atomic systems, including both lifetime- and Doppler-broadened media. These activation functions serve as core components for implementing AONNs. Compared with earlier approaches based on two-level saturable absorbers~\cite{10.1117/1.AP.7.1.016004}, our scheme offers significantly greater tunability and reconfigurability. Most notably, we propose and experimentally demonstrate two-channel nonlinear optical activation functions for the first time.

These dual-channel functions enable more expressive transformations while preserving energy efficiency and bandwidth—surpassing the capabilities of conventional electronic and hybrid photonic-electronic architectures. Our results confirm that atomic vapor media offer a compact, tunable, and reconfigurable platform for implementing scalable nonlinear activation. Furthermore, their compatibility with integrated photonic systems positions them as a promising candidate for edge computing and real-time optical processing.

\vspace{0.5cm}
\noindent\textbf{Outlook.} 
Multichannel activation functions, especially those with multiple-input multiple-output (MIMO) capabilities, provide significant advantages by capturing correlations between different input and output signals. This expressiveness is valuable in a wide range of domains. In image super-resolution, MIMO activations enable convolutional neural networks to process color channels jointly, yielding more accurate reconstructions~\cite{Wang2019}. In denoising autoencoders, they improve robustness by modeling inter-channel correlations~\cite{Zhang2017}. In natural language processing and transformer models, MIMO activations support simultaneous processing of multiple embeddings, enhancing contextual understanding~\cite{Ramachandran2017}. In robotics, autonomous vehicles, and IoT systems, where multimodal sensor fusion is critical, MIMO activations help capture relationships between signals from diverse sources~\cite{Huang2020}.

Figure~\ref{fig:fig11} illustrates candidate AONN architectures. Figures~\ref{fig:fig11}(a) and (b) show traditional single-channel implementations based on sigmoid and ReLU nonlinearities, respectively. Figure~\ref{fig:fig11}(c) introduces a MIMO AONN enabled by two-channel nonlinear optical activation. This multichannel architecture reduces the required network depth and expands the repertoire of nonlinear operations while maintaining the energy efficiency of quantum-interference-based mechanisms.

Experimental results from Figs.~\ref{fig:fig02}--\ref{fig:fig09} indicate that the activation functions typically operate at optical powers on the order of 100~$\mu$W. Assuming 10\% loss per layer, this power level can support up to seven sequential neurons, resulting in approximately 17~$\mu$W per neuron. Extrapolating this to a full network, one could support one million optical neurons with only 17~W of total optical power.

Despite this promise, challenges remain—particularly in scaling to larger networks, managing thermal effects, and ensuring stability and noise resilience. Continued progress in photonic integration, quantum materials, and system-level co-design will be critical to overcoming these limitations. As AI workloads increasingly demand high-speed, low-power computation, AONNs based on atomic vapor or atom-like materials such as quantum dots and solid-state defects represent a compelling path forward. This study lays the foundation for future exploration of optimized activation designs, alternative materials, and integrated system implementations that will unlock the full potential of optical neural computing.

\begin{acknowledgments}
The work was supported by Purdue University startup and NSF Grant No. 2211989.
\end{acknowledgments}

\vspace{0.5cm}
\noindent\textbf{Code and Data Availability.} 
\noindent\ Code for numerical simulations and experimental data are available at: \url{https://github.com/AndersonXu99/AONN-Nonlinearity}

\vspace{0.5cm} \noindent\textbf{Competing financial interests}\\ 
\noindent The authors declare the following competing interests: S.D., H.A., R.C. and
X.X. are listed as inventors on US provisional patent application 63/781,585 filed by Purdue University.
All remaining authors declare no competing interests.

\appendix


\section{Numerical Simulation} \label{Appendix: Numerical Simulation}

Simulations were conducted in Python using NumPy and SciPy to solve the steady-state density matrix equations describing the three-level atomic system. The density matrix elements were computed through direct matrix inversion, $x = A^{-1}b$, under steady-state conditions. To incorporate Doppler broadening, the atomic response was integrated over a Maxwell–Boltzmann velocity distribution using numerical quadrature. Systematic parameter sweeps of laser power, detuning, and temperature were used to characterize nonlinear optical responses. Simulation accuracy was validated by comparing with analytical limits and experimental data from warm vapor measurements.

\section{Experimental Details} \label{Appendix: Experimental Details}
Experiments were performed using a 2-cm-long cylindrical $^{87}$Rb vapor cell with anti-reflection coating for the D1 transition. The cell contained no buffer gas and was housed in thermal insulation without magnetic shielding. Two optical fields—either collinear or counter-propagating—were aligned through the cell as shown in Fig.~\ref{fig:fig01}(b), with orthogonal linear polarizations for efficient filtering. Both beams were focused using a 2-f system with a 250 mm lens, achieving a $1/e^2$ waist of 75~$\mu$m and a Rayleigh range of approximately 4 cm.

Input-output nonlinear activation measurements were obtained by modulating laser power via a piezo-controlled waveplate and a polarizing beam splitter. Probe power fluctuations were kept below 5\%, even in the low-power regime.

System response time was characterized using pulsed inputs for the probe field, generated via a digital delay/pulse generator (SRS DG645) and an electro-optic modulator (Eospace AZ-0K5-10-PFU-PFU-795-UL). The transmitted signal was detected by a photomultiplier tube (Hamamatsu H10721-20) and recorded on a digital oscilloscope (Tektronix TDS684B, 1 GHz, 5 GS/s). All detection hardware had rise times $\leq$2~ns.




\bibliography{AONN}





\end{document}